\documentclass[useAMS,usenatbib,onecolumn]{mn2e}
\usepackage{graphicx}
\usepackage{bm}
\usepackage{epsfig}
\usepackage{amssymb, ulem}
\usepackage{amsmath}
\usepackage{empheq}
\usepackage{amsfonts}
\usepackage{cleveref}
\usepackage{bm}
\usepackage{booktabs}
\usepackage{tabularx}
\newcommand{\bea}{\begin{eqnarray}}
\newcommand{\eea}{\end{eqnarray}}

\newcommand{\beq}{\begin{equation}}
\newcommand{\eeq}{\end{equation}}

\newcommand{\simless}[0]{\mathbin{\lower 3pt\hbox
   {$\rlap{\raise 5pt\hbox{$\char'074$}}\mathchar"7218$}}}
\newcommand{\simgreat}[0]{\mathbin{\lower 3pt\hbox
   {$\rlap{\raise 5pt\hbox{$\char'076$}}\mathchar"7218$}}}

\newcommand{\figref}[1]{figure \ref{#1}}

\newcommand{\capfigref}[1]{Figure \ref{#1}}

\newcommand{\eqnrefs}[1]{eqs. (\ref{#1})} %ditto
\title[Axis ratio evolution from phase space dynamics]{Evolution of axis ratios from phase space dynamics of triaxial collapse}
\author[Sharvari Nadkarni-Ghosh and Bhaskar Arya]{Sharvari Nadkarni-Ghosh$^{1}$\thanks{E-mail:
sharvari@iitk.ac.in, nsharvari@gmail.com} and Bhaskar Arya $^{2}$\thanks{E-mail:
bhaskararya4@gmail.com} \\
$^{1}$Department of Physics, I.I.T. Kanpur, Kanpur, U.P. 208016 India \\
$^{2}$Department of Physics, I.I.T. Kanpur, Kanpur, U.P. 208016 India}

\begin{document}
\date{}
\pagerange{\pageref{firstpage}--\pageref{lastpage}} \pubyear{}

\maketitle
\label{firstpage}
\begin{abstract}
We investigate the evolution of axis ratios of triaxial haloes using the phase space description of triaxial collapse. In this formulation, the evolution of the triaxial ellipsoid is described in terms of the dynamics of eigenvalues of three important tensors: the Hessian of the gravitational potential, the tensor of velocity derivatives and the deformation tensor. The eigenvalues of the deformation tensor are directly related to the parameters that describe triaxiality, namely, the minor to major and intermediate to major axes ratios ($s$ and $q$) and the triaxiality parameter $T$. Using the phase space equations, we evolve the 
eigenvalues and examine the evolution of the PDF (probability distribution function) of the axes ratios as a function of mass scale and redshift for Gaussian initial conditions. We find that the ellipticity and prolateness increase with decreasing mass scale and decreasing redshift. These trends agree with previous analytic studies but differ from numerical simulations. However, the PDF of the scaled parameter ${\tilde q}  = (q-s)/(1-s)$ follows a universal distribution over two decades in mass range and redshifts which is in qualitative agreement with the universality for conditional PDF reported in simulations. We further show using the phase space dynamics that, in fact, ${\tilde q}$ is a phase space invariant and is conserved individually for each halo. These results, demonstrate that the phase space analysis is a useful tool that provides a different perspective on the evolution of perturbations and can be applied to more sophisticated models in the future.

\end{abstract}

\maketitle

\section{Introduction}

Observations have established beyond any doubt that dark matter haloes are well described by a triaxial geometry. The evidence has come from a variety of probes including the optical and X-ray surface brightness, indirect observations through the Sunyaev-Zeldovich effect and weak and strong gravitational lensing (see review by \citealt{limousin_three-dimensional_2013} and references therein). Numerical simulations have also confirmed triaxiality
(e.g.,  \citealt{jing_triaxial_2002,allgood_shape_2006,schneider_shapes_2012}) and ellipsoidal halo-shape finders have proven to give more realistic haloes than their spherical counterparts \citep*{despali_ellipsoidal_2013}. 
The motivation for understanding the shape, structure and dynamics of haloes is two-fold. In the currently accepted hierarchical model of structure formation small scale structures collapse first and larger structures are formed through mass accretion and major mergers. The details of the halo shape distribution provide useful information regarding this process of structure formation and evolution. The second major reason is from the point of view of precision cosmology. For example, understanding the systematics due to intrinsic ellipticity correlations is a crucial part of any weak lensing measurement (e.g., \citealt{refregier_weak_2003,oguri_effects_2004,joachimi_intrinsic_2013}) and important in determining the Hubble constant from cluster shapes \citep{wang_systematic_2006,kawahara_systematic_2008}.
Cluster and void ellipticities have also been proposed as a probe to constrain cosmological parameters such as $\sigma_8$ and the dark energy equation of state \citep{ho_cluster_2006,lee_axis-ratio_2006,park_void_2007,lavaux_precision_2010}. 

Over the years, numerical simulations have become increasingly sophisticated and efficient at the task of modelling dark matter halo shapes (\citealt{dubinski_structure_1991,bullock_shapes_2002,jing_triaxial_2002,kasun_shapes_2005}; \citealt*{hopkins_cluster_2005}; \citealt{allgood_shape_2006}; \citealt*{schneider_shapes_2012,despali_like_2014}; \citealt{bonamigo_universality_2015}; \citealt{suto_evolution_2016}; \citealt*{vega-ferrero_shape_2017}). Using this information for precision cosmology involves exploring a large range of parameter space and the computational cost involved in N-body codes proves to be a significant drawback. Furthermore, as was pointed out by \citet{allgood_shape_2006}, various groups differ in their methodology of determining halo shapes and hence differ in their results. Thus, the need for analytic investigations  still remains. 

Triaxial collapse in an expanding universe has been under study for over four decades \citep{icke_formation_1973,white_growth_1979, nariai_dynamics_1972, barrow_growth_1981,eisenstein_analytical_1995, bond_peak-patch_1996, lithwick_self-similar_2011}. Various authors have considered it specifically in the context of axis ratio evolution. The seminal paper by \cite{bardeen_statistics_1986} used excursion set theory to compute the ellipticity and prolateness distribution of dark matter haloes forming under gravitational collapse of initial Gaussian random fields. \citet*{lee_analytic_2005} combined the dynamics given by the Zeldovich approximation with the statistical properties of the linear field to obtain the probability distributions of axis ratios. \citet{sandvik_why_2007} and \cite{desjacques_environmental_2008} have used excursion sets to model the effect of the environment on halo ellipticity. 

Recent work by \citet{nadkarni-ghosh_phase_2016}, hereafter NS16, analysed triaxial collapse in terms of the coupled dynamics of three tensors: the Hessian of the gravitational potential , the tensor of derivatives of the velocity field and the deformation tensor, whose eigenvalues are denoted as $\lambda_d, \lambda_v$ and $\lambda_a$ respectively. The dynamical equations governing these eigenvalues were derived from the commonly used \cite{bond_peak-patch_1996} model for triaxial collapse. This reformulation serves two purposes: first, it gets rid of the complicated elliptic integrals in the original formulation and second, it provides a natural way to track the dynamics of the various perturbation fields. Examining the nine-dimensional `phase space of the resulting dynamical system allows us to view the evolution of perturbations from a different perspective. In particular, finding invariant subsets in phase space can help in constructing or interpret relations between the variables that are valid throughout the evolution. The main focus of NS16 was to understand the relation between the gravitational and velocity fields and their joint evolution in the non-linear regime. By examining the dynamical behaviour in phase space, the authors were able to find a new universal relation between $\lambda_d$s and $\lambda_v$s that was valid over a wide range of mass scales and redshift. In this paper, we focus on the dynamics of the deformation tensor, which gives direct information about the evolution of the axis ratios. Recently, this dynamical systems approach has been used in other contexts as well: \citet{nadkarni-ghosh_non-linear_2013} used it to compute the non-linear density velocity divergence relation based on spherical collapse and \cite{nadkarni-ghosh_einstein-boltzmann_2017} used it to gain insights into the Einstein-Boltzmann system which governs the evolution of cosmological perturbations at early epochs. 

The paper is organized as follows. In \S \ref{setup}, we review the phase space equations and establish the notation. \S \ref{runs} describes the details of the numerical runs. The shape of the triaxial object is quantified using three parameters: the minor to major axis ratio, $s$, the intermediate to major axis ratio, $q$, and the triaxiality parameter $T$ \citep*{franx_ordered_1991}. These can be defined in terms of the eigenvalues of the deformation tensor whose dynamics is completely determined by the phase space equations. \S \ref{results} gives the results and we discuss and conclude in \S \ref{conclusion}.

\section{The phase space equations}
\label{setup}
The physical system consists of a homogenous isolated ellipse evolving in a cosmological background consisting only of dark matter and dark energy with an equation of state $w=-1$ ($\Lambda$CDM). In the absence of rotations, the evolution of the ellipse is completely determined by the three tensors: the Hessian of the gravitational potential, the tensor of velocity derivatives and the deformation tensor. Let $\lambda_{d,i}, \lambda_{v,i}$ and $\lambda_{a,i}$ \{i=1,2,3\}, denote the eigenvalues of these three tensors respectively. Let $a$ be the scale factor of the background and $a_i$ be the scale factors corresponding to the three axes of the ellipsoid. Assuming that the principle axes of all three tensors are the same throughout the evolution, the $\lambda_i$s and $a_i$s are related as 
\begin{subequations}
\label{params}
\begin{align} 
\label{ladef}
\lambda_{a,i} &= 1- \frac{a_i}{a} \;\;\;\;\; \\
\label{lvdef}
\lambda_{v,i} &= \frac{1}{H} \frac{{\dot a}_i}{a_i} -1 \\ 
\label{lddef}
\lambda_{d,i} &= \frac{\delta \alpha_i}{2} + \lambda_{ext,i}, 
 \end{align}
 \end{subequations}
 where
 \bea
\label{alphadef} 
\alpha_i &=& a_1a_2a_3 \int_0^\infty \frac{d \tau}{ (a_i^2 + \tau) \prod_{j=1}^{j=3}(a_j^2 + \tau)^{1/2}}  \;\; \; \; \mbox{with} \; \left(\sum_{i=1}^3 \alpha_i=2\right)\\
\lambda_{ext,i} &=& \frac{5}{4} \left(\alpha_i - \frac{2}{3} \right) \; \; \; \; \mbox{non-linear approx}. 
\label{lambdaext}
\eea
The \citet{bond_peak-patch_1996} model of triaxial collapse gives evolution equations for the three axis lengths $a_i$s. These can be rewritten as evolution equations for the three sets of eigenvalues (see NS16 for details). The equations are:  
\begin{subequations}
\label{phspdyn}
\begin{align}
\label{dyn1}\frac{d \lambda_{a,i}}{d \ln a} &= -\lambda_{v,i}(1-\lambda_{a,i})\\
\label{dyn2}\frac{ d \lambda_{v,i}}{d \ln a}& = -\frac{1}{2} \left[ 3 \Omega_m(a) \lambda_{d,i}  -  \left\{ \Omega_m(a) - 2 \Omega_\Lambda(a)-2 \right \}  \lambda_{v,i}   + 2 \lambda_{v,i}^2\right] \\
\label{dyn3}\frac{d \lambda_{d,i}}{d \ln a} &= -(1+\delta) \left(\delta + \frac{5}{2} \right)^{-1} \left(\lambda_{d,i} + \frac{5}{6} \right)  \sum_{j=1}^3\lambda_{v,j} \\
\nonumber & + \left( \lambda_{d,i} + \frac{5}{6} \right) \sum_{i=1}^3 (1+\lambda_{v,i})  - \left( \delta + \frac{5}{2} \right)(1+\lambda_{v,i}) \\
\nonumber & + \sum_{j\neq i}\frac{ \left\{\lambda_{d,j} - \lambda_{d,i} \right\} \cdot \left\{(1-\lambda_{a,i})^2(1+\lambda_{v,i}) -(1-\lambda_{a,j})^2(1+\lambda_{v,j})\right\}}{(1-\lambda_{a,i})^2-(1-\lambda_{a,j})^2},
\end{align}
\end{subequations}
where
\beq
\nonumber \delta = \sum_{i=1}^3 \lambda_{d,i}.
\eeq
and the evolution of $\Omega$s is given by 
 \beq 
 \Omega_{m}(a)  = \frac{ \Omega_{m,0}  H_0^2a_0^3}{H^2 a^3};  \; \; \; \Omega_\Lambda(a) = 1-\Omega_{m}(a) \label{omega}
 \eeq
 The subscripts `0' denotes the value today. Note that this reformulation is devoid of the complicated elliptic integrals that occur in the original formulation. These get absorbed in the definition of the $\lambda_{d,i}$. At early times, when the fluctuations are small, $\lambda_{d,i} = \lambda_{a,i}$ and $\lambda_{v,i} = - f(\Omega_m) \lambda_{d,i}$, where $f(\Omega_m)$ is the linear growth rate, usually approximated as $f(\Omega_m) \approx \Omega_m^{0.55}$ for a $\Lambda$CDM cosmology \citep{linder_cosmic_2005}. The initial conditions are specified at $a= a_{init}$. These are described in the next section. 

\section{Numerical Runs}
\label{runs}
\begin{figure}
\includegraphics[height=6cm]{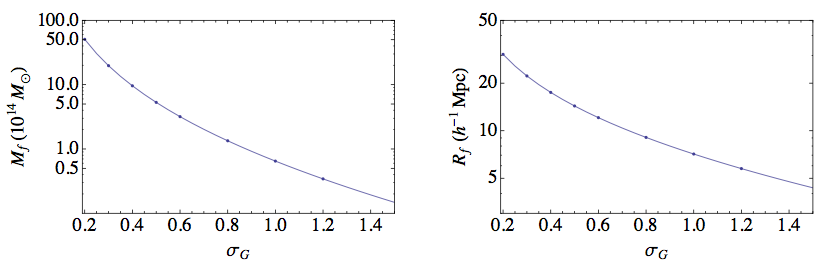}
\caption{Relation between the r.m.s. $\sigma_G$ and the mass scale $M_f$ for a BBKS power spectrum with spectral index $n_s = 1$ and a Gaussian filter $W(k R_f)  = e^{-\frac{k^2 R_f^2}{2}}$. The power spectrum was normalized such that $\sigma_8 = 0.9$. The points indicate the $\sigma_G$ values chosen for the analysis. } 
\label{msig}
\end{figure}

 For Gaussian initial conditions, the distribution of the $\lambda_{d,i}$ initially is given by \cite{doroshkevich_spatial_1970} 
\beq 
 p(\lambda_{d,1}, \lambda_{d,2}, \lambda_{d,3})  = \frac{15^3}{8 \pi \sqrt{5} \sigma_G^6} \exp \left(-\frac{3I_1^2}{\sigma_G^2} + \frac{15 I_2}{2 \sigma_G^2}\right) \times (\lambda_{d,1} -\lambda_{d,2})(\lambda_{d,2} -\lambda_{d,3})(\lambda_{d,1} - \lambda_{d,3})
\label{pdf}
\eeq
where $\sigma_G$ is the r.m.s. fluctuation at the scale $R_f$, $I_1 = \lambda_{d,1} + \lambda_{d,2} + \lambda_{d,3}$ and $I_2 = \lambda_{d,1} \lambda_{d,2} +  \lambda_{d,2} \lambda_{d,3} +  \lambda_{d,1} \lambda_{d,3}$. This distribution only assumes Gaussianity and does not depend on the details of the power spectrum. The power spectrum is relevant when relating $\sigma_G$ to a mass or radius scale:  
\beq 
\sigma_G^2(R_f,z)  = \frac{1}{(2\pi)^3} \int P(k,z) W^2(k R_f) d^3k, 
\eeq
where $P(k,z)$ is the linear power spectrum at redshift $z$ and depends on the power spectrum at redshift zero through the linear growth factor \citep{dodelson} and $W(k R_f)$ is the window function. The mass scale $M_f$ is related to smoothing scale $R_f$ as $M_f = 4 \pi/3 R_f^3 {\bar \rho}_m$, where ${\bar \rho}_m$ is the homogenous background matter density. \capfigref{msig} shows the relation between $\sigma_G$ and the mass and radius scales for the BBKS power spectrum \citep{bardeen_statistics_1986, dodelson} with a Gaussian window function. We choose the cosmological parameters in accordance with the $\Lambda$CDM cosmology dictated by WMAP-7 \citep{komatsu_seven-year_2011}: $\Omega_{m,0} = 0.29, \Omega_{\Lambda,0} = 0.71, n_s = 1, \sigma_8 = 0.9, h = 0.73$. The critical density of the universe is ${\rho}_{c,0} = 2.775 h^2 \times 10^{11} M_{\odot} {\rm Mpc}^{-3}$ \citep{dodelson}. 

We consider eight values of $\sigma_G$ in the range $\sigma_G = 0.2-1.2$. This covers almost two decades in mass range, from about $5 \times 10^{15} M_{\odot}$ to $3 \times 10^{13} M_{\odot}$.  For every value of $\sigma_G$ we draw five realizations, each consisting of $5\times10^4$ points. As a proxy to simulate haloes, we restrict to only those points for which all three initial eigenvalues are positive (these correspond to density peaks)\footnote{In NS16, the authors showed that the predictions of the non-linear density-velocity divergence relation based on ellipsoidal collapse agreed better with simulations, when the initial conditions were restricted to density peaks, which are characterized by all three initial eigenvalues positive (see figure 6).}. This corresponds to only eight percent of the total sample (see fig. C1 in appendix of NS16). At the initial time $a_{init}$, chosen to be $0.001$, the system is linear and $\Omega_m \approx 1$, implying $\lambda_{a,i}(a_{init}) = \lambda_{d,i}(a_{init}) $, $\lambda_{v,i}(a_{init}) = -\lambda_{d,i}(a_{init})$. The initial conditions were evolved from $a_{init}$ to $a=1$ using the system of \eqnrefs{phspdyn}. The shape parameters of the ellipse at any intermediate time can be constructed from the $\lambda_{a}$s as outlined below. 

The shape of a triaxial object is completely characterized by two parameters (a third parameter will correspond to setting the scale of the object). It is customary to use the ratios of smallest to largest and intermediate to largest axes defined as 
\bea 
s &=& \frac{a_{min}}{a_{max}}\\
q&=& \frac{a_{inter}}{a_{max}}, 
\eea
where $a_{min}, a_{max}$ and $a_{inter}$ are the smallest, largest and intermediate axis respectively. $s=0$ when $a_{min}=0$ i.e., the axis has collapsed. In terms of the $\lambda_a$ parameters, these are 
\bea 
s &=& \frac{1-\lambda_{a,max}}{1-\lambda_{a,min}}\\
q&=& \frac{1-\lambda_{a,inter}}{1-\lambda_{a,min}}. 
\eea
The triaxiality parameter that characterizes the prolateness is defined as \citep*{franx_ordered_1991} 
\beq 
T = \frac{a_{max}^2-a_{inter}^2}{a_{max}^2-a_{min}^2} = \frac{1-q^2}{1-s^2}.
\eeq
Note that $T$ is not an independent parameter, but derived from $q$ and $s$. Often, in the literature, triaxiality is characterized in terms of the ellipticity $e$ and prolateness $p$ \citep{bardeen_statistics_1986} and haloes are known to populate a triangular region in the $e-p$ space \citep*{porciani_testing_2002, desjacques_environmental_2008,despali_like_2014}. These are equivalent variables, defined from the eigenvalues of the deformation tensor. While both sets are easy to measure in simulations, $s$ and $q$ have a more direct geometric interpretation. In this paper, we examine the evolution of $s$, $q$ and $T$ as a function of mass and redshift for the $\Lambda$CDM cosmology described above. For any epoch, we analyse only those haloes for which $s, q >0$, i.e., the halo has not collapsed. Here we focus only on dark matter overdensities; the investigation of voids is left for future studies.

\section{Results}

\label{results}
\begin{figure}
\includegraphics[height=6cm]{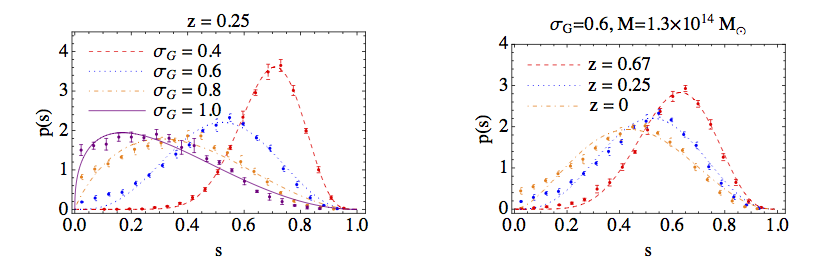}
\caption{Probability density function of the minor to major axis ratio. The raw PDF is fit by a beta distribution. Smaller mass haloes (larger $\sigma_G$) are more elliptical (smaller $s$) than the larger mass haloes. For a fixed mass scale, the halo ellipticity increases at lower redshifts.   }
\label{sratio}
\end{figure}
\capfigref{sratio} shows the probability distribution function (PDF) of $s$. The left panel shows the variation with mass scale at a given redshift and right panel shows the variation with redshift for a fixed mass. Smaller mass haloes (larger $\sigma_G$) are more elliptical (smaller $s$) than the larger mass haloes. For a fixed mass scale, the halo ellipticity increases at lower redshifts. These trends are in agreement with the results of \cite*{lee_analytic_2005} who used the Zeldovich approximation to analytically model the dynamics and connect the initial and final axis ratios. 

However, these conclusions are contrary to the trends observed in numerical simulations. Multiple numerical simulations (e.g.,\mbox{\citealt{jing_triaxial_2002, allgood_shape_2006,bonamigo_universality_2015, suto_evolution_2016, vega-ferrero_shape_2017}}) have shown that lower mass haloes are more spherical and the average ellipticity for a fixed mass scale decreases at lower redshifts. One obvious possibility for this difference is that the analytic model considered here deals with isolated ellipsoids; the effect of the environment, accretion and merging are not incorporated. These factors play an important role in determining halo ellipticity (see for e.g., \citealt{jing_triaxial_2002}). Low mass haloes collapse earlier and become more spherical due to accretion and mergers. Therefore, at a given redshift, the lower mass haloes are more spherical than higher mass haloes. Similarly, haloes of similar mass are more elliptical at a higher redshift since mergers are fewer at higher redshifts. Another important source of discrepancy is the assumption of homogenous evolution inbuilt in the analytic model. In the Bond and Myers model of ellipsoidal collapse, the the forces on the ellipsoid are taken to be linear in the coordinates. As a result, the collapse is self-similar, i.e., the ellipsoid remains an ellipsoid and the density inside remains homogenous (see \citealt{eisenstein_analytical_1995}).  However, recent work by \cite{suto_evolution_2016} has shown that the internal density distribution is inhomogeneous and the shape of a halo is far from self-similar. In particular, ellipsoids fitted to the inner and outer regions of haloes do not have the same orientation; inner ellipsoids are also more elongated that outer ones indicating that the axis ratio is also different. These authors found that the predictions from the homogenous ellipsoidal collapse agreed well with simulations only until the turn-around epoch.  

In fact, ellipticity trends predicted by the analytic models can partly be traced to the initial conditions. For a fixed mass scale, the uniform ellipsoidal collapse simply amplifies the initial ellipticity and hence for any halo, mean ellipticity increases with decreasing redshift. Trends with mass scale for a given redshift can be understood based on Bernardeau's analytic results for the evolution of rare peaks. \cite{bernardeau_nonlinear_1994} showed that the non-linear evolution of a rare peak is well described by spherical collapse. Given a fixed initial $\delta$, a larger mass implies a larger length scale and a smaller $\sigma_G$. That is, the initial peak height ($\nu = \delta/\sigma_G$) is larger, implying a rarer peak. For any epoch, including the initial one, the rarer the peak, the more spherical it is. Hence smaller $\sigma_G$ (more massive) haloes tend to be more spherical and less ellipsoidal.

Recent numerical simulations by \cite{bonamigo_universality_2015} and \cite{vega-ferrero_shape_2017} have found a universal form for the pdf 
when the variable $s$ is scaled by factor proportional to a power of the critical peak height $\nu = \delta_c/\sigma(M)$, where $\delta_c$ is the critical overdensity and $\sigma(M)$ is the variance of the initial density field smoothed on a mass scale $M$ (referred to as $\sigma_G$ in this paper). We do not reproduce such a universal scaling for $s$. Instead we find that $s$ is is well fit by a $\beta$ distribution given by the general form 
\beq 
p(x,\alpha, \beta) = \frac{1}{B(\alpha, \beta)} x^{\alpha-1} (1-{\tilde x})^{\beta-1},
\eeq
where $B(\alpha, \beta)$ is the normalizing factor given by $B(\alpha, \beta) = \int_0^1 x^{\alpha-1} (1-x)^{\beta-1} dx$. 
 The parameters $\alpha$ and $\beta$ depend on mass scale and redshift and we fit them as functions of $\sigma_G$ and $\Omega_m$ are given in appendix \S \ref{fits}.  
 Earlier studies based on simulations \citep{oguri_arc_2003,suto_evolution_2016} have also found that the projected axis ratio follows a beta distribution. We do not expect that our fitting functions will match any actual data, but they may serve as guess functions for a more accurate modelling in the future.

\begin{figure}
\includegraphics[height=6cm]{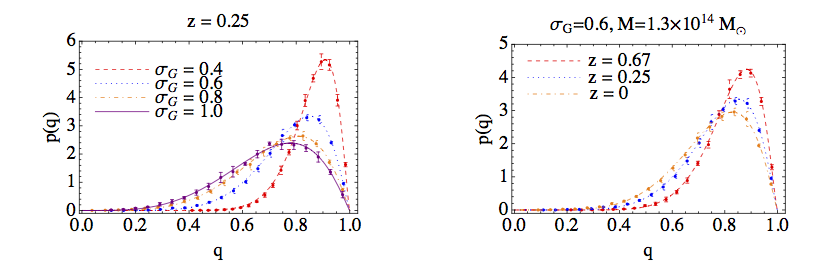}
\caption{Probability density function of the intermediate to major axis ratio. The trends are the same as those captured by the $s$-ratio; ellipticity increases with decreasing redshift and decreasing mass scale. The raw PDFs are fit by a beta distribution. }
\label{qratio}
\end{figure}

\begin{figure}
\includegraphics[height=6cm]{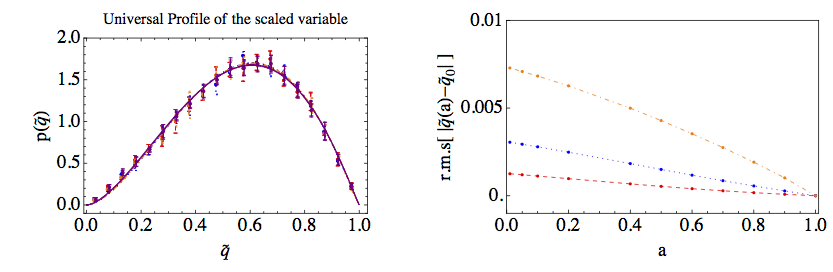}
\caption{${\tilde q}$ as an invariant of the dynamics. The left panel shows the PDF of ${\tilde q}$ for the four $\sigma_G$ values and three redshift values considered in \figref{qratio}. The PDFs of the scaled variables overlap indicating a universality. The right panel shows the difference between ${\tilde q}$ and its value today for three values of $\sigma_G$ shown by the red dashed ($\sigma_G = 0.2$), blue dotted ($\sigma_G = 0.4$) and orange dot-dashed ($\sigma_G = 1.2$) lines. The difference stays small throughout the evolution indicating that at early times ${\tilde q}$ was close to its value today. Thus illustrates that ${\tilde q}$ may be considered as an invariant quantity. }
\label{error}
\end{figure}

\capfigref{qratio} shows the PDF of intermediate to major axis ratio. The left panel shows the variation with respect to mass at a fixed redshift and and the right panel shows the variation with redshift at fixed mass.The qualitative trends are the same as those inferred from the $s$ ratio.  The ratios $s$ and $q$ individually capture the two dimensional eccentricity along a cross-section perpendicular to the intermediate and minor axis respectively. To understand the full 3D geometry, it is generally the conditional probability $p(q|s)$ which is the quantity of interest. \cite*{schneider_shapes_2012} found that 
the conditional probabilities were largely independent of mass and redshift. They proposed the transformation to the variable to the $s$-dependent variable 
\beq 
{\tilde q} = \frac{q-s}{1-s}.
\eeq
Their motivation for this transformation was that its range in the unit interval (because $q<s$ and $q<1$) and hence it has the same support as the beta distribution. Redshift independent conditional probabilities have also been reported recently by \cite{bonamigo_universality_2015} based on analysis of the MXXL and SBARBINE simulations \citep{angulo_scaling_2012,despali_universality_2016} and by \cite*{vega-ferrero_shape_2017} based on the MultiDark simulations \footnote{http://www.multidark.org, https://www.cosmosim.org}. We do not consider the conditional probability $p({\tilde q}|s)$ i.e., we do not bin in $s$ space. Instead, we use the $q$ and $s$ values of the individual density peak (halo) and find that the PDF of this transformed variable is well fit by a 
universal beta distribution with parameters $\alpha = 2.68$ an $\beta = 2.07$ over the mass range $\sigma_G = 0.2 -1.2$ and over the redshift range $z=99$ to $z=0$.

The agreement at very early times, suggests that this scaling is encoded in the initial distribution, but the universality suggests that the non-linear evolution preserves it. To understand this universality better, we examine the evolution of ${\tilde q}$ for a individual perturbation over the entire redshift range. For each point in the realization we compute the difference $|{\tilde q}(a)-{\tilde q}_0|$, where ${\tilde q}_0 = {\tilde q}(a=1)$ is the transformed ratio today and find that it decreases with time. The right panel of \figref{error} shows the root mean square difference $\langle |{\tilde q}(a)-{\tilde q}_0| \rangle$ (taken over five realizations)\footnote{We also checked this for the maximum absolute error for a single realization and found that it also decreases monotonically.} for three values of $\sigma_G$ shown by the red dashed ($\sigma_G = 0.2$), blue dotted ($\sigma_G = 0.4$) and orange dot-dashed ($\sigma_G = 1.2$) lines. It is clear that the difference stays small throughout the evolution and we conclude that to within about a 1\% error, ${\tilde q}$ can be thought of as a invariant of the dynamics. The fact that the difference decreases monotonically is just a consequence of defining the differences with respect to the value today (where, by definition, the error is zero). 

Incidentally, it is also possible to construct the variable ${\tilde q}$ for the density and velocity sectors: 
\bea 
{\tilde q}_d &=& \frac{q_d-s_d}{1-s_d} = \frac{\lambda_{d,max} - \lambda_{d,inter}}{\lambda_{d,max} - \lambda_{d,min}},\\
{\tilde q}_v &=& \frac{q_v-s_v}{1-s_v} = \frac{\lambda_{v,max} - \lambda_{v,inter}}{\lambda_{v,max} - \lambda_{v,min}}.
\eea
In NS16, the authors found that $q_d$ and $q_v$ are also related by a universal relation $q_d + q_v  =1$, which holds to percent level accuracy over all redshifts and a similar range of mass scales studied here.

\begin{figure}
\includegraphics[height=6cm]{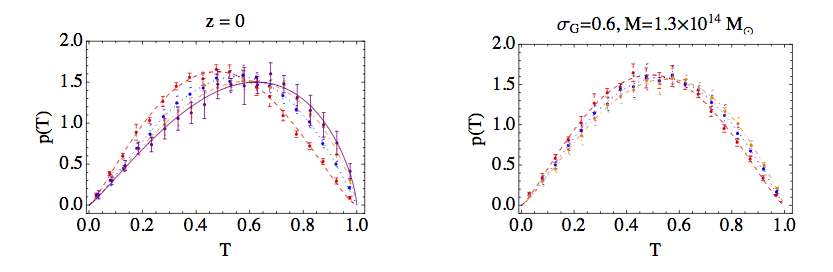}
\caption{Probability density function of the triaxiality parameter $T$. The colour coding is same as that of \figref{sratio}. The tendency for prolateness increases with decreasing redshift and decreasing mass.}
\label{Tratio}
\end{figure}

\capfigref{Tratio} shows the distribution of the triaxiality parameter $T$. A perfectly prolate spheroid is one in which two axes are equal and less than the third ($a_{min} = a_{inter} < a_{max} \implies q=s, T=1$) 
and a perfectly oblate spheroid is one where two axes are equal and greater than the third ($a_{min}<a_{inter} = a_{max} \implies q=1, s<1, T=0$). For general triaxial objects with no two axes equal, the range $0<T<1/3$ is considered oblate, $2/3<T<1$ is considered prolate. Haloes with $1/3<T<2/3$ are neither prolate nor oblate (\citealt{allgood_shape_2006}; \citealt*{schneider_shapes_2012}).  The left panel of \figref{Tratio} shows that at a given redshift, lower mass haloes (higher $\sigma_G$) are more prolate than higher masses and the right panel shows that 
prolateness is higher for lower redshifts. On the other hand, numerical simulations (\citealt*{schneider_shapes_2012,despali_like_2014}) and observations \citep{paz_alignments_2011} suggest that most haloes are prolate, the more massive haloes being more prolate. Prolateness also increases with increasing redshift. These contradicting results again highlight the importance of post collapse processes, accounted for in simulations, in determining halo prolateness.

\section{Discussion and Conclusion}
\label{conclusion}
In this paper, we have investigated the evolution of halo axis ratios using the phase space description of triaxial collapse, recently developed by \cite{nadkarni-ghosh_phase_2016}. In this method, the triaxial collapse is completely described by the joint evolution of the eigenvalues of three tensors: the Hessian of the gravitational potential, the tensor of velocity derivatives and the deformation tensor. The equations describing this nine dimensional `phase-space' are derived from the Bond and Myers model of triaxial collapse \citep{bond_peak-patch_1996}. The phase space reformulation has the advantage that it gets rid of the elliptic integrals that occur in the original system and is a natural framework to study the joint dynamics of the three tensors that govern the evolution of the triaxial halo. Moreover, it is a better way to study universal relations between the variables since they can be interpreted as phase space invariants. 

Three parameters quantify the shape of a triaxial object: the ratio of minor to major axis $s$, the ratio of intermediate to major axis $q$, and the triaxiality parameter $T$, which quantifies prolateness or oblateness. These are directly related to the eigenvalues of the deformation tensor. We evolved the phase space equations using initial conditions drawn from a Gaussian distribution characterized by the width $\sigma_G$. We chose eight values of $\sigma_G$ that spanned over decade in mass scale from roughly $0.5 -50 \times 10^{14} M_{\odot}$. We computed the PDFs of $s$, $q$ and $T$ as a function of redshift and found that for a given redshift lower mass haloes are more elliptical and more prolate than higher mass haloes and the ellipticity and prolateness decreases with increasing redshift. These trends are opposite to those predicted in simulations, but in agreement with other analytic studies such as \cite{lee_analytic_2005}. We found that the PDFs of $s$, $q$ and $T$ are well fit by a beta distribution and fit the parameters of the distribution as functions of redshift as mass. Recent numerical simulations \citep{bonamigo_universality_2015, vega-ferrero_shape_2017}
have reported a universal PDF for a scaled $s$ and also a universal behaviour in the conditional probability for the transformed variable ${\tilde q} = (q-s)/(1-s)$. We do not find universality in either $s$ or $q$ directly, but find that the transformed variable ${\tilde q}$ has indeed a universal PDF. We further examine the dynamics in phase space to illustrate that, in fact, ${\tilde q}$ is an invariant of the dynamics and is preserved for each halo individually. Similar phase space invariants have been studied in the context of density-velocity dynamics (\citealt{nadkarni-ghosh_non-linear_2013, nadkarni-ghosh_phase_2016}). If such invariants exists, then they can potentially be exploited to obtain additional information that is not directly measurable or place additional constraints on cosmological parameters.

The trends for the axial ratio evolution obtained here are opposite to those from simulations. This lack of agreement suggests that the analytical model needs revision. 
The dynamical evolution equations used in this work were derived from the triaxial collapse model of of \cite{bond_peak-patch_1996}. This model involves many approximations. The initial configuration is a uniform density ellipsoid which undergoes uniform collapse. The equations of motion do not involve rotation and the effect of external tidal forces is modelled in terms of the axes of the ellipsoid. It is worth exploring modifications which may give more realistic predictions. 
A non-uniform density configuration or non-uniform collapse could be potential extensions. Alternately, one may have to consider models which more accurately takes into account the effect of the environment \citep{desjacques_environmental_2008} and/or employ a more general formalism that allows for rotations (for example \citealt{nariai_dynamics_1972,barrow_gravitational_1993,eisenstein_analytical_1995}). Another treatment of homogenous mathematical models in the context of general relativity can also be found in \cite{bogoyavlenskii_homogeneous_1976}. The main motivation to build increasingly accurate analytic models, in spite of the increasing accuracy of N-body simulations, comes from the point of view of precision cosmology. Given the range of unconstrained cosmological parameters and the increasing size of data sets, computing time becomes a limiting factor. Analytic alternatives are hence necessary and studies like the one presented here pave the way for more sophisticated model building and analysis in the future. 

\section{Acknowledgements}
SN would like to acknowledge the Science and Engineering Research Board (SERB), Dept. of Science and Technology (DST India) for the research grant (YSS/2014/000526). SN would also like to thank Akshat Singhal for discussions during the initial stages of this project. 

\appendix
\section{Fitting functions for the beta distributions}
\label{fits}
We found that the PDFs  for the $s$, $q$ and $T$ variables are well described by beta distributions with parameters $\alpha$ and $\beta$ given by 
\beq 
p(x,\alpha, \beta) = \frac{1}{B(\alpha, \beta)} x^{\alpha-1} (1-{\tilde q})^{\beta-1},
\eeq
where $B(\alpha, \beta)$ is the normalizing factor given by $B(\alpha, \beta) = \int_0^1 x^{\alpha-1} (1-{\tilde q})^{\beta-1} dx$. 
 The parameters $\alpha$ and $\beta$ depend on mass scale and redshift. For each variable, $s$, $q$ and $T$, we fit these parameters as functions of $\Omega_m(z)$ and $\sigma_G$, where $\Omega_m(z)$ is given by 
 \beq
 \Omega_m(z) = \frac{H_0^2 \Omega_{m,0} (1+z)^3}{H^2(z)}.
 \eeq
 We denote the parameters with appropriate subscripts $s,q$ and $T$ depending upon the variable. The fitting forms are
\begin{subequations}
\begin{align}
\log_{10} \alpha_s &= A_s(\sigma_G) \Omega_m(z) + B_s(\sigma_G)\\
\log_{10} \beta_s &= C_s(\sigma_G) \Omega_m(z) + D_s(\sigma_G)\\
A_s(\sigma_G) &=0.23 + 1.55 \sigma_G\\
B_s(\sigma_G) &= 1.95 -3.08 \sigma_G\\ 
C_s(\sigma_G) &= -0.11 + 0.87 \sigma_G\\
D_s(\sigma_G) &= 1.03 -1.01 \sigma_G
\end{align}
\end{subequations}

\begin{subequations}
\begin{align}
\log_{10} \alpha_q &= A_q(\sigma_G) \Omega_m(z) + B_q(\sigma_G)\\\
\log_{10} \beta_q &= C_q(\sigma_G) \Omega_m(z) + D_q(\sigma_G)\\
A_q(\sigma_G) &= 0.35 + 0.54 \sigma_G\\
B_q(\sigma_G) &= 1.74 -2.04 \sigma_G\\ 
C_q(\sigma_G) &= -0.03 + 0.23 \sigma_G\\
D_q(\sigma_G) &= 0.47 -0.31 \sigma_G
\end{align}
\end{subequations}
Fitting forms for both the $s$ and $q$ variables has an error of 25\% over the mass range $\sigma =0.3-0.6$ and epoch range $a=0.6-1$. 
\begin{subequations}
\begin{align}
\alpha_T &= 2.17\\
\log_{10} \beta_T &= C_T(\sigma_G) \Omega_m(z) + D_T(\sigma_G)\\
C_T(\sigma_G)  &=  -0.05 + 0.33\sigma_G  \\
D_T(\sigma_G) &=  0.48 -0.34 \sigma_G
\end{align} 
\end{subequations}
The fitting form for the $T$ parameter has an error of about 20\% over the mass range $\sigma = 0.2 -1.2$ and epoch range $a=0.4-1$ and less than 10\% over the range 
$\sigma =0.3-0.6$ and $a=0.6-1$.

%\clearpage
%\section{References}
%The style file is journal specific. 
\bibliographystyle{mn2e.bst}
\bibliography{ellbibtex1,ellbibtex2,ellbibtex3}

\end{document}